\documentclass[sigconf]{acmart}
\AtBeginDocument{%
  }

\setcopyright{acmlicensed}
\copyrightyear{2018}
\acmYear{2018}
\acmDOI{XXXXXXX.XXXXXXX}
\acmConference[Conference acronym 'XX]{FoSE}
\acmISBN{978-1-4503-XXXX-X/2018/06}

\usepackage{csquotes}
\newcommand*{\enq}[1]{\enquote{{\itshape#1}}}

\usepackage{balance}

\begin{document}

\title{Ethics of Care for Software Engineering}


\author{Alexander Serebrenik}
\email{a.serebrenik@tue.nl}
\orcid{0000-0002-1418-0095}
\affiliation{%
  \institution{Eindhoven University of Technology}
  \country{The Netherlands}
}

\author{Sebastian Baltes}
\email{sebastian.baltes@uni-heidelberg.de}
\orcid{0000-0002-2442-7522}
\affiliation{%
  \institution{Heidelberg University}
  \country{Germany}
}


\begin{abstract}
Software engineering researchers repeatedly argue that the impact of their research on industrial practice, while desired and intended, is rarely achieved.
We believe that a possible explanation of this phenomenon is the opposition of ``caring about'' and ``caring for'', based on the \emph{ethics of care}.
Indeed, while software engineering is collaborative and hence builds on interpersonal relations, researchers tend to care about ``industrial impact'' and ``practitioners'' in abstract terms, but rarely care for specific individuals working in specific contexts facing specific challenges.
In this position paper, we advocate for the adoption of ethics of care in software engineering and discuss the implications of this adoption for researchers and conference organizers.
\end{abstract}


\maketitle

A recent survey of software engineering researchers conducted by Margaret-Anne Storey and Andre van der Hoek as part of the Future of Software Engineering track~\cite{storey2025communitySurvey} surfaced multiple concerns about the divergence between academic research in software engineering and industrial practice.
When asked what aspect or aspects of the software engineering research community do not work well and why, respondents lamented \enq{minimal impact on industry and almost no relevance to adjacent fields like AI}, \enq{lack of industrial relevance}, the research community being too \enq{far from real issues and problems found by practitioners}, and engaging in \enq{too much navel gazing <...> and  not enough focused on translating results to industry and impact}.
This concern is not unique to the survey respondents. 
\enq{Software engineering research has had as much impact on programmers as astronomy has had on stars,} as Greg Wilson has aptly put it.\footnote{https://third-bit.com/talks/to-dont/\#3}
However, at the same time, survey respondents regret that, compared to other computer science researchers, software engineering researchers are \enq{often perceived as less relevant, less impactful, and overly focused on industrial application} and remind that as researchers \enq{we are not competing with industry, we are doing something different than them, and we often forget this.}
The fundamental question is the role of the industry vs. academia relation in software engineering.
To what extent is engagement with the industry fundamental for our discipline or is it merely the way it is currently being organized?

A possible answer is that, in return for rewarding careers, good salaries, and public recognition, we as academics have the duty to generate useful knowledge for the respective state and to train a workforce capable of applying it~\cite{Richardson2025Entities}.
Given that most universities are paid by the government, a close connection to the needs of the state, which is paying the salaries and other expenses might, however, threaten the validity of scientific results.
In extreme cases, this might lead to the suppression of research areas, for example, the suppression of genetics in the Soviet Union and the creation of pseudoscience such as Lysenkoism~\cite{harding1991whose}. 

Another angle is that (applied) research can be judged on its application in practice.
This is particularly true in software engineering, where the usefulness of knowledge becomes equated with the impact on industrial practice. 
However, while science can be \enq{the pacemaker of technological progress}~\cite{Bush1945EndlessFrontier}, it has also been argued that the value of a scientific inquiry cannot be reduced to the set of prospective yet uncertain technological applications (cf.\ discussion of purism by Kitcher~\cite{kitcher2001science}).

However, Sommerville, in his classic software engineering textbook, argues that \enq{software engineering is intended to support professional software development, rather than individual programming}~\cite{sommerville2011software}.
This means that software engineering as a discipline can only exist in relation to the practice of software development.
As this practice is carried out by people, we interpret Sommerville's definition as a recognition that \textbf{software engineering as a discipline is defined through its relation with and for people}.
The ethics of care considers relationships not from a legalistic point of view but as having an intrinsic unconditional and non-negotiable value~\cite{starratt1991bulding}.
This, together with our interpretation of software engineering as a discipline, means that \textbf{software engineering should be subject to the ethics of care}~\cite{gilligan1993different,noddings2003caring}.

The ethics of care is based on three ideas~\cite{russell2019make}. 
First, people are fundamentally dependent on each other. 
Empirical software engineering researchers inherently depend on practitioners, as researchers study the process enacted by practitioners or artifacts created by them.
Second, the ethics of care calls for attention to the most vulnerable when making decisions. 
There is increasing attention to making software development more diverse and inclusive~\cite{hyrynsalmi2025making}. 
Finally, the ethics of care differentiates between a more abstract notion of ``caring about'' and more immediate ``caring for''~\cite{noddings2003caring}, and argues that our moral choices should attend and respond to the immediate conditions of our context.

It is particularly the last opposition between ``caring about'' and ``caring for'' contains promise of resolving the opposition between software engineering researchers trying to impact the industry but reaching a limited success. Therefore, we pose the following question: 

\begin{quote}
    \emph{Do software engineering researchers really ``care for'' software practitioners, that is, specific individuals developing software or do they rather ``care about'' practical impact on an abstract level?}
\end{quote}

This opposition calls for the replacement of generic ``developers'', ``companies'', and ``projects'' that we care about in our research with specific developers, companies, and projects that we care for.
Trying to impact software development practice in all its diversity might not be feasible; however, we can and should aim to make software development a better place for individuals.


In addition, individual researchers should continue to work with people from vulnerable or underprivileged populations within software engineering.
Recent studies have considered the experiences of software developers from minoritized groups such as women~\cite{cutrupi2026gender}, neurodiverse developers~\cite{gama2025stgt,newman2025get}, racial and ethnic minority developers~\cite{dagan2023builing}, LGBTIQ+ developers~\cite{santos2023benefits}, older developers~\cite{baltes2020is}, developers with disabilities~\cite{saben2024enabling}, developers from the Global South~\cite{ragkhitwetsagul2025impact} and intersections of these identities~\cite{breukelen2023still,szlavi2024designing}.
This list is by no means exhaustive and should not be limited to demographic diversity.
Perspectives of developers from small companies that do not necessarily have direct connections to universities or academic research, of scientists trained outside of computer science and developing software to support their research, or end-users creating websites with Generative AI should be explored and included in our view of software engineering.
The ``caring for'' principle requires deep and prolonged engagement with those populations, and acknowledgment of the right of these individuals to be who they are~\cite{starratt1991bulding}. 

Similarly, conference organizers should prioritize topics that affect practitioners as opposed to those that simply improve technology, and studies conducted in a specific context and having had a positive impact in this context.
\textbf{We call for each and every software engineering paper to present a story of a real person} who develops software (whether seeing themselves as a software engineer or not), or a person who teaches or researches software development.
This presentation should include a description of their specific context and the specific challenges they face, e.g., related to the technology they are using, their way of working, or their workplace environment.
Such a story---anonymized or not---can strengthen our connection to practice, answer the fundamental question why are we doing software engineering research (because we care), and hopefully help to achieve the impact sought after by the survey respondents.

\section*{Acknowledgments}
The authors thank Mary Shaw and Titus Winters for sharing their thoughts on the topic on the Discord channel.

\balance
\bibliographystyle{ACM-Reference-Format}
\bibliography{software}
\end{document}